\documentclass[aip,pop,amsmath,amssymb,reprint,superscriptaddress,longbibliography]{revtex4-1}

\usepackage{graphicx}
\usepackage{array}[=2016-10-06]
\usepackage{dcolumn}
\usepackage{bm}
\usepackage[utf8]{inputenc}
\usepackage[T1]{fontenc}
\usepackage{mathptmx}
\usepackage{mathtools}
\usepackage{textcomp}
\newlength{\figurewidth}
\usepackage{etoolbox}
\usepackage{xurl}
\usepackage[hidelinks]{hyperref}
\hypersetup{pdftitle={Multiple Instance Learning for Interpretable Plasmoid Detection in the FLARE Device},pdfauthor={Charles Marsden, Adam Robbins, Hantao Ji}}

\makeatletter
\def\@email#1#2{%
 \endgroup
 \patchcmd{\titleblock@produce}
  {\frontmatter@RRAPformat}
  {\frontmatter@RRAPformat{\produce@RRAP{*#1\href{mailto:#2}{#2}}}\frontmatter@RRAPformat}
  {}{}
}%
\makeatother

\begin{document}


\title{Multiple Instance Learning for Interpretable Plasmoid Detection in the FLARE Device} 



\author{Charles Marsden}
\email{cmarsden@uchicago.edu}
\affiliation{Princeton Plasma Physics Laboratory, Princeton, NJ, USA}
\affiliation{University of Chicago, Chicago, IL, USA}
\author{Adam Robbins}
\affiliation{Princeton Plasma Physics Laboratory, Princeton, NJ, USA}
\affiliation{Princeton University, Princeton, NJ, USA}
\author{Hantao Ji}
\affiliation{Princeton Plasma Physics Laboratory, Princeton, NJ, USA}
\affiliation{Princeton University, Princeton, NJ, USA}



\begin{abstract}
Multiple instance learning (MIL) is a form of weakly supervised learning widely used in healthcare and biomedical imaging, where labels may be available for groups of observations (``bags") but not individual observations (``instances"). This setup is well-suited to plasmoid detection in magnetic reconnection experiments, since identifying dynamic magnetic structures is best accomplished taking in data from many time steps as a whole, as opposed to viewing frames in isolation.  Instance-level consideration enables researchers to determine the range of times most affecting model output, thus serving as a tool for the temporal localization of plasmoids. Despite its success in the life sciences, to the best of our knowledge, MIL is yet to be reported in peer-reviewed plasma physics literature. We evaluate a MIL model against a convolutional recurrent neural network (CRNN), an architecture commonly found in plasma physics literature, on a plasmoid detection task, using a highly imbalanced dataset of 1152 magnetic reconnection shots from the FLARE device at PPPL. We demonstrate significant improvements in test performance (PR AUC: $\Delta=+0.0932$, $p=0.0039$; F1: $\Delta=+0.0259$, $p=0.0488$; Wilcoxon signed-rank test) against the CRNN model on full-channel input. We investigate the interpretability advantage provided by our MIL model for the temporal localization of plasmoids. Through its attention mechanism, we can discern relevant frames and even the number of plasmoids present in a shot. Overall, we show that MIL is a promising framework for similar event classification tasks in plasma physics.
\end{abstract}


\maketitle 

\section{\label{sec:intro}Introduction}
The Facility for Laboratory Reconnection Experiments (FLARE) was proposed as a next-generation device for studying the physics of magnetic reconnection and related phenomena \cite{ji2026flarefacility}. Its goal is to provide access to reconnection physics at high Lundquist numbers and large normalized system sizes, allowing scientists to study plasmas in regimes relevant to those in astrophysical systems.

Reconnection is a fundamental plasma process in which the rearrangement of magnetic field topology converts magnetic field energy into kinetic energy of particles \cite{Loureiro_2016}. It is an important phenomenon for astrophysical systems ranging from solar flares and pulsar winds to molecular clouds and black hole accretion disks \cite{JiPhaseDiagram}. Plasmoids are magnetic islands of closed flux formed from reconnecting current sheets. The breakup of current sheets due to the tearing instability and the subsequent formation of plasmoids appear to be general features of large reconnecting systems \cite{Loureiro2007}. Current sheets in astrophysical plasmas frequently exhibit Lundquist numbers on the order of $S\sim 10^{12}$--$10^{20}$ $~$\cite{JiPhaseDiagram}. Numerical simulations suggest that current sheets with Lundquist numbers of $S \gtrsim 10^4$ are tearing-unstable and thus break up into chains of plasmoids \cite{JiPhaseDiagram}. Therefore, plasmoids are likely to be abundant in astrophysical reconnecting systems. 

FLARE has recently demonstrated its ability to access reconnection regimes favorable to plasmoid formation \cite{ji2026flarefacility}. With certain favorable machine parameters, up to nearly half of shots contain plasmoids. While, in principle, finding a closed flux surface is necessary and sufficient evidence for the presence of a plasmoid, data shortcomings including noise and limited spatial resolution mean that plasmoids must be identified heuristically to rule out false positives. The following question arises naturally: how can we accelerate the identification of plasmoids in FLARE? One such answer is to implement machine learning methods.

Previous attempts at predictive modeling with machine learning methods have made extensive use of recurrent neural networks (RNNs). These networks are designed to process sequential data by maintaining a hidden state that is combined with each new input to generate new outputs. \citet{KatesHarbeck2019} implemented an RNN variant known as a long short-term memory (LSTM) model for tokamak disruption prediction. They trained their model to output a time-dependent prediction of disruption probability, showing strong performance on JET and DIII-D data. Importantly, time-resolved labels could be constructed for the training data because the disruption times were known. Similarly, \citet{Matos2020} used a CRNN to classify low, dithering, and high confinement states in the TCV tokamak. Their recurrent CRNN model outperformed a CNN for confinement state classification, demonstrating the importance of long-term temporal dependencies in plasma diagnostics. A subsequent sequence-to-sequence model with attention developed by \citet{Matos2021} used a ConvLSTM with an autoregressive decoder for confinement state classification, achieving substantial improvement on an already strong baseline ConvLSTM model. Labels were determined by the mutual consent of two experts for each time step in each series. They paid great attention to short-term confinement modes on the order of 2 ms, with an inherent label determination uncertainty on the order of 1 ms, highlighting the labeling burden associated with strongly supervised methods for plasma time series classification.


Success with reduced labeling has been demonstrated in a plasma physics context. \citet{Montes2021} applied semi-supervised label spreading to detect events in tokamak discharges with few manually labeled shots, achieving $>85\%$ detection accuracy for DIII-D H--L back transitions and initially rotating locked modes. \citet{Zheng2020} developed a disruption predictor for J-TEXT using a CNN-LSTM architecture trained only on non-disruptive discharges. The model learned to recognize differences between predicted and observed signals as an anomaly indicator for impending disruption events.

Non-recurrent models have also found success in plasma physics settings. \citet{Churchill2020} implemented a temporal convolutional network (TCN) for time series classification and disruption prediction in fusion devices. Their TCN contained four hidden layers with dilated convolutions of different dilation factors, allowing the model to capture both short and long-term dependencies. They labeled frames in the final 300 ms before the disruption time as `disruptive', achieving an F1 score of $91\%$ across individual time frame predictions. \citet{Bergstedt2024} developed a supervised CNN for detecting plasmoids in spacecraft measurement-like simulation data. The authors used particle-in-cell current sheet simulations to generate data then derive plasmoid locations from the magnetic topology. They obtained labels with high spatial resolution, using plasma bulk velocity in addition to the magnetic probe and current density data used for our work as the five input channels. Their proof-of-concept classifier achieved a recall of 0.79 and a precision of 0.40 for an F1 score of $\sim 0.53$ on held-out simulation data. They emphasized that good performance on simulated data does not guarantee similar performance on experimental data. Their work 
highlights the challenge of obtaining reliable labels and the broader difficulty of automated plasmoid detection itself.

In this study, we systematically investigate the capabilities of a multiple instance learning model against a CNN-LSTM model. We investigate whether gated-attention MIL can support the interpretation and temporal localization of plasmoid signatures. We show that, through its attention mechanism, we can discern the extent to which individual frames contribute to shot-level series representations, determine times coinciding with plasmid-like structure, and identify the number of distinct plasmoids present in a shot. Comparing classification performance, the MIL model exhibits statistically significant and substantial improvements relative to the CRNN with the full input data. However, the results also indicate that more work is needed to develop the model before it can be widely adopted in plasma physics.

\section{Models}
Certain architectural components are common to both convolutional recurrent neural networks (CRNNs) and multiple instance learning (MIL) models. Convolutional neural networks (CNNs) are used as the feature extractor in both models. Additionally, both approaches seek to aggregate information from frame-level vectors into a single shot-level vector representation. The final shot classifier for each model is a single-hidden-layer multi layer perception mapping input vectors to a probability in $(0,1)$.


\subsection{Multiple instance learning (MIL)}
In the standard binary classification setting one aims to approximate some function mapping a given input feature vector $\mathbf{x} \in \mathbb{R}^D$ to $\{0, 1\}$. In the MIL setting, such vectors are known as instances, and we deal with ``bags" that are collections of instances $\mathbf{X}_i = \{\mathbf{x}_{i,1}, \dots, \mathbf{x}_{i,T}\} \in \mathcal{X} = \{\mathbf{X}_{1}, \dots, \mathbf{X}_{n}\}$ $~$\cite{Doran2016}. There exists a known bag-labeling function $F : \mathcal{X} \rightarrow \{0,1\}$ such that $Y_i = F(\mathbf{X}_i)$, determined via manual bag labeling, and for a given bag $\mathbf{X}_i$ there exists an unknown instance-labeling function $f : \mathbb{R}^D \rightarrow \{0,1\}$ such that $y_{i,t}=f(\mathbf{x}_{i,t})$ for $t\in\{1,\dots,T\}$. Thus we have a set of known binary bag labels $Y = \{Y_1,\dots,Y_n\}$ and for a given bag $\mathbf{X}_i$ an associated unknown set of binary instance labels $\{y_{i,1},\dots,y_{i,T}\}$. The bag-labeling function itself is the piecewise-defined \cite{Dietterich1997}
\begin{equation}
Y_i = 
    \begin{cases}
    0, & \text{iff } \sum_t y_{i,t}=0 \\
    1, & \text{otherwise.}
    \end{cases}
\end{equation}
Alternatively, we can write
\begin{equation}
    Y_i=\max_t\{y_{i,t}\}.
\end{equation}


\subsubsection{Gated attention and pooling}
The attention mechanism involves linearly projecting each instance into independent attention spaces of dimension $d'=\max(32,\lfloor{d/2}\rfloor)=64$, then calculating scalar attention scores for each frame and using these scores to combine the frame representations into a single vector. With parameters $\mathbf{U},\mathbf{V} \in \mathbb{R}^{64 \times 128}$, $\mathbf{w} \in \mathbb{R}^{64 \times 1}$, the scalar attention score for frame $\mathbf{x}_{i,t}$ is calculated  \cite{Ilse2018}
\begin{equation}\label{eq:attention_score}
    a_{i,t} = \frac{\exp\{\mathbf{w}^\mathrm{T}(\tanh(\mathbf{V}\mathbf{x}_{i,t}^\mathrm{T}) \odot \sigma( \mathbf{U}\mathbf{x}_{i,t}^\mathrm{T}))\}}
    {\sum_{j=1}^{T} \exp\{\mathbf{w}^\mathrm{T} (\tanh(\mathbf{V}\mathbf{x}_{i,j}^\mathrm{T}) \odot \sigma( \mathbf{U}\mathbf{x}_{i,j}^\mathrm{T}))\}}
\end{equation}
where $\odot$ denotes the Hadamard (element-wise) product. The $\tanh$ activation function forces elements of the vector $\mathbf{V}\mathbf{x}_{i,t}^\mathrm{T}$ to be in $(-1,1)$. The sigmoid function $\sigma$
forces elements of the vector $\mathbf{U}\mathbf{x}_{i,t}^\mathrm{T}$ to be in (0,1). Qualitatively, the $\mathbf{V}$ branch provides a learned representation of the content of each frame, while the $\mathbf{U}$ branch acts as a learned gate that filters the information flowing through it. 
The shot-level representation is calculated with a weighted sum of the frame representations. For a bag $\mathbf{X}_i$, we have
\begin{equation}\label{eq:MIL_pooling}
    \mathbf{z}_i = \sum_{t=1}^{80} a_{i,t} \mathbf{x}_{i,t} \in \mathbb{R}^{128}
\end{equation}
where $\sum_{t=1}^{80}a_{i,t} = 1$. Note that the aggregation operator is permutation invariant; the order of frames in each time series has no effect on model outcomes.

\subsection{Recurrent neural networks (RNNs)\label{sec:RNNs}}
RNNs process time series data sequentially, maintaining a hidden state vector that is a function of all previous inputs. As such, they can natively handle long-term patterns in time. A mathematical treatment of CRNNs can be found in Ref.~\onlinecite{Schmidt2019}. We use a long short-term memory (LSTM) variant of the CRNN for our model comparison.


\section{Methodology}

\subsection{Dataset}
Our dataset consists of 1152 experimental shots from the FLARE device from March through May 2026. For details of the FLARE facility and data collection, see Ref.~\onlinecite{ji2026flarefacility}. Each shot consists of 80 time-ordered 2-dimensional spatial maps in the same R--Z reconnection plane of the FLARE device at 1 \textmu s intervals from times 261--340 \textmu s of the shot cycle. The data at the same point in all spatial profile arrays correspond to the same physical coordinate in this reconnection plane. A shot is a ``plasmoid shot" if a plasmoid was identified at some point in the shot's run (every plasmoid naturally occurred within the specified time interval). In total, there are 168 plasmoid shots ($14.58\%$) in the dataset. For training purposes, the data was split into a train-validation set of 979 shots of which 143 ($14.60\%$) are plasmoid shots and a held-out test set of 173 shots of which 25 ($14.45\%$) are plasmoid shots. Clearly, the class imbalance is significant.

We considered four possible 2D profile physical quantities for the models: BR, BZ, JY and BY. BR and BZ are the two in-plane magnetic field components, BZ being the reconnecting component and BR the perpendicular component for the pull reconnection considered \cite{ji2026flarefacility}. Together, these components carry information about the reconnecting magnetic field topology. They contain in-plane field structure and thus encode information regarding the location and geometry of the current sheet, X- and O-points, and the motions of plasmoids. Since plasmoids are effectively a structural feature of the magnetic field topology, these two channels are of high importance.

\subsection{Preprocessing}\label{sec:Preprocessing}
To train the model, we manually classified each shot according to the likelihood that a plasmoid was formed from the current sheet due to the plasmoid instability. In principle, one can simply classify shots algorithmically by whether or not they contain closed isocontours of the flux $\psi = \int 2 \pi R B_Z \; dR - 2 \pi R B_R \; dZ$. In practice, however, this naive approach generates many false positives for several reasons. Firstly, the inherent random measurement error in our magnetic field measurement and processing often results in small closed flux surfaces without any physical underpinning. Such ``structures" tend to either remain suspiciously stationary or vanish rapidly. Secondly, processes other than the plasmoid instability can create genuine closed-flux structures. Most common is the case of pinch-offs from the flux core (see \citet{gerhardt2008} for details). These can be easily distinguished by eye due to their great size and rapid movement from outside the magnetic probes but have identical magnetic geometries to plasmoids. 

While many cases are clear-cut, others, especially involving small potential plasmoids, remain uncertain even after careful inspection. Thus we label FLARE shots for plasmoid presence according to a confidence scale from 0 to 3.  Shots with no evidence of plasmoids were given a score of `0'. Shots with some evidence for plasmoids, but still less likely than not, were given a score of `1'. Shots with inconclusive evidence, but more likely than not, were given a score of `2'. Shots with definitive evidence for plasmoids were given a score of `3'. Note that since shots commonly contain more than one plasmoid, only the plasmoid with the greatest certainty is considered. The highest confidence shots contain plasmoids which manifestly originate from the current sheet, persist over time, and display physical growth and ejection behavior. The importance of plasmoid dynamics in identification is a chief reason it is well-suited to MIL as opposed to methods treating frames in isolation.

All time series receive binary labels indicating whether or not a plasmoid was present. We implement two separate label policies in this work. The first policy excludes low confidence `1' shots, keeping `0' shots as the negative class (non-plasmoid shots) with `2' or `3' shots kept as the positive class (plasmoid shots). In total, this set contains 1152 shots. This policy was used for all but one trial. The second policy includes 73 additional `1' shots in the positive class.

All shots of the training set were Z-score normalized to a per-channel mean of 0 and standard deviation of 1. Validation and test set shots are normalized with the training set mean and standard deviation.

We generated two augmented sets of JY data to see if augmentation improved performance. The first was created by reflecting the spatial profiles in the $Z = 0$ axis. The second set was created by adding Gaussian noise drawn from $\mathcal{N}(0,0.05\sigma)$ where $\sigma$ is that spatial profile's standard deviation. When using augmented sets, we also applied online augmentation on every new sample in each new training epoch by applying additional Gaussian noise from $\mathcal{N}(0,0.01)$.





\subsection{Hyperparameter search}


We conducted a hyperparameter search through 24 combinations of 12 variables affecting the MIL model architecture. 
Each cross-validation 3-fold used $85\% \times \frac{2}{3}=56.7\%$ of the full dataset for training and $85\% \times \frac{1}{3}=28.3\%$ for validation and early-stopping.
We used the AdamW optimizer throughout. The variables we tuned, as well as the searched combinations (where applicable) and sampling strategy can be found in the GitHub repository linked under ``Data Availability". The per-fold probability threshold for a plasmoid vs non-plasmoid shot was determined by the value which maximized the F1 score. The optimal parameters were a plain, (16, 32, 64) CNN architecture, embedding dimension of 128, dropout of 0.25, learning rate of $1\times10^{-4}$, weight decay of $1\times10^{-3}$, batch size 8, BCE loss, cosine scheduler, classifier multiplier of 1.0, and online noise augmentation of 0.01$\sigma$.

The optimal model achieved an F1 score of $0.590 \pm 0.075$, PR AUC of $0.513 \pm 0.149$, and ROC AUC of $0.858 \pm 0.048$. 
The search landscape was flat near the optimum with no one model as a standout; only learning rate had a notable non-random effect on model performance. Thus the selected configuration should be regarded as one representative of a broad number of near-equivalent models rather than as the optimal configuration.


\subsection{MNIST-Bags}
To verify that our model is successful outside of our current scope of plasmoid classification, we tested our model on an MNIST classification task \cite{LeCun1998, Deng2012} adapted for multiple instance learning models. The model achieved an F1 score of 0.934, PR AUC of 0.983, ROC AUC of 0.979, precision of 0.947, and recall of 0.922. These results are broadly consistent with those found in published literature \cite{Ilse2018}. This serves as verification that our implementation produces expected behavior on a MIL-specific dataset.



\section{Model testing}\label{sec:Model_testing}
Throughout this stage of testing, we used a configuration found through previous exploration that differs from the optimized only in dropout (0.3 vs. 0.25) and excluding online augmentation. Its cross-validation F1 (0.573 $\pm$ 0.053) is well within one standard deviation of the selected configuration (0.590 $\pm$ 0.075), with per-fold paired differences of 0.003–0.042. We used this configuration for the MIL model and CRNN. They shared training hyperparameters (such as learning rate, encoder width, and batch size) while architecture parameters do not deviate from other common architectures (such as embedding dimension, number of recurrent layers, unidirectionality). We found that the differences in performance between the MIL and CRNN models with full-dataset input are much greater than any configuration-to-configuration variability observed in the hyperparameter search. Thus, for the CRNN model with the same CNN frame encoder as the MIL model, followed by a single-layer unidirectional LSTM with hidden dimension 256, the results of a CRNN-specific hyperparameter search would likely have been negligible.

We examined the effects of input channels (JY, BR–BZ, and BR–BZ–BY–JY), compared MIL and CRNN models using JY and full inputs, the effects of excluding low-confidence shots, and applying offline augmentation using JY alone. Note that all test data and results can be found in the GitHub repository linked under ``Data Availability Statement". We split the train-validation data with a different random seed to the fixed seed used for the train-validation/test split. Single-exclusion comparisons allowed the effects of individual components to be isolated. We repeated this process for ten different random seeds (101--110). Using a different seed after initial test set separation enables us to account for seed-to-seed model performance variability; affected components of the training process are: the three cross-validation folds, train/validation split, random weight initialization, and data loader ordering. We recorded the F1, PR AUC, ROC AUC test scores, as well as the 3-fold cross-validation mean F1 score. For each comparison we thus had ten paired data points per metric, one from each seed's trial, and we could then conduct statistical tests on these paired samples.

We conducted Wilcoxon signed-rank tests across the ten seeds. Paired Student's $t$-tests were additionally conducted, but the sample size of $n=10$ seeds does not guarantee that differences are normally distributed so the Wilcoxon signed-rank test is likely more reliable. We used Welch's $t$-test for testing our labeling policy. All tests were two-sided.

\section{Results}
\begin{table*}[t]
\caption{Performance of each experimental trial over ten training seeds.
Values are the mean $\pm$ standard deviation across ten seeds. Except where
indicated, all trials use the confidence-2/3 label policy with 979
train-validation shots and 173 held-out test shots.}
\label{tab:per_trial_results}
\centering
\small
\begin{ruledtabular}
\begin{tabular}{@{}p{0.31\linewidth}cccc@{}}
Experimental trial
    & Test F1
    & Test PR AUC
    & Test ROC AUC
    & Mean CV F1 \\
\hline
Gated-attention, JY
    & $0.528 \pm 0.052$
    & $0.537 \pm 0.036$
    & $0.805 \pm 0.018$
    & $0.565 \pm 0.012$ \\

Gated-attention, BR--BZ
    & $0.581 \pm 0.045$
    & $0.635 \pm 0.062$
    & $0.820 \pm 0.016$
    & $0.588 \pm 0.012$ \\

Gated-attention, BR--BZ--BY--JY
    & $0.578 \pm 0.024$
    & $0.633 \pm 0.023$
    & $0.836 \pm 0.012$
    & $0.585 \pm 0.013$ \\

CRNN, JY
    & $0.531 \pm 0.029$
    & $0.502 \pm 0.066$
    & $0.813 \pm 0.020$
    & $0.525 \pm 0.010$ \\

CRNN, BR--BZ--BY-JY
    & $0.552 \pm 0.049$
    & $0.540 \pm 0.053$
    & $0.835 \pm 0.016$
    & $0.531 \pm 0.017$ \\

Gated-attention, JY, augmentation off
    & $0.527 \pm 0.041$
    & $0.556 \pm 0.026$
    & $0.810 \pm 0.010$
    & $0.563 \pm 0.012$ \\

Gated-attention, JY, augmentation on
    & $0.538 \pm 0.056$
    & $0.572 \pm 0.042$
    & $0.815 \pm 0.017$
    & $0.561 \pm 0.009$ \\

Gated-attention, JY, confidence 1 included\textsuperscript{a)}
    & $0.566 \pm 0.021$
    & $0.582 \pm 0.026$
    & $0.799 \pm 0.007$
    & $0.594 \pm 0.008$ \\
\end{tabular}
\end{ruledtabular}
\par\smallskip
\parbox{\linewidth}{\raggedright\footnotesize\textsuperscript{a)}\,The confidence-1 trial uses a different and larger dataset:
1041 train-validation shots and 184 held-out test shots. Its metrics are
therefore not directly paired with those of the other trials.}
\end{table*}
The results of the Wilcoxon signed-rank tests are presented in this section. Other test results can be found at Ref.~\onlinecite{marsden2026zenodo}. Here, ``JY baseline" refers to the MIL model discussed in section \ref{sec:Model_testing} with JY data as the sole input. The full per-trial results, with means and standard deviations across the ten seeds for each of the four metrics, are shown in table \ref{tab:per_trial_results}.

\subsection{Channels}


BR–BZ and the full BR-BZ-BY-JY input produced nearly identical F1 and PR AUC. Adding BY and JY to BR–BZ did not measurably improve either metric ($\Delta < 0.0015; p > 0.84$). Including all channels improved mean ROC AUC by 0.017. It also had substantially lower seed-to-seed variability in F1 and PR AUC. Therefore, the four-channel model appears to provide more stable ranking performance, but BR and BZ alone captured most of the practically relevant F1 and PR AUC improvement. This result is encouraging for 2-dimensional reconnection models as the inclusion of JY and BY data did not substantially improve the model's test performance.

\subsection{Architecture}
We found that, on JY data alone, the CRNN and MIL models perform similarly on the held-out test set. The superior performance of the MIL model as measured by PR AUC ($+0.0351$, $p=0.084$) is suggestive but not significant at the conventional $p=0.05$ threshold. However, the MIL model scored higher on mean CV F1 ($+0.040$, $p=0.00195$) and outperformed the CRNN on every seed ($W=0$). This suggests that the MIL model was more consistently effective across folds during training, but the test set was too small or variable to demonstrate a held-out test advantage.

\begin{table}[t]
\caption{Wilcoxon signed-rank comparison of the gated-attention MIL model against the CRNN using four-channel BR--BZ--BY--JY
input.}
\label{tab:wilcoxon_crnn_vs_gated_four_channel}
\centering
\small
\begin{ruledtabular}
\begin{tabular}{@{}lccc@{}}
Metric & Mean difference & $W$ & $p$ \\
\hline
Test F1
    & $+0.0259$ & 8 & $0.048828$ \\
Test PR AUC
    & $+0.0932$ & 1 & $0.003906$ \\
Test ROC AUC
    & $+0.0016$ & 27 & $1.000000$ \\
Mean CV F1
    & $+0.0535$ & 0 & $0.001953$ \\
\end{tabular}
\end{ruledtabular}
\end{table}
The gated-attention MIL model significantly outperformed the CRNN across test F1, PR AUC, and CV F1 with the full BR--BZ--BY--JY input (see table \ref{tab:wilcoxon_crnn_vs_gated_four_channel}). The MIL model had substantially greater PR AUC ($+0.09$) and mean CV F1 ($+0.05$) and both differences were highly consistent across seeds ($W=1$, $W=0$, respectively). Test ROC AUC was effectively identical ($+0.0016$, $W=27$, $p=1.000$) indicating that the two architectures had similar global positive-negative ranking performance. However, improved model performance with similar ROC AUC scores can occur when large class imbalances are present in the dataset. The significantly improved PR AUC of the MIL model indicates that it had greater ability to accurately identify members of the minority class. The MIL model exhibits a nominally significant improvement ($p=0.048$), but we did not include any multiple-comparison corrections and the standard Bonferroni correction would likely scale the $p$ value to greater than $0.05$. It also achieved a precision of $0.525 \pm 0.061$, a recall of $0.656 \pm 0.051$, and an accuracy of $86.1 \pm 2.1 \%$. These are promising considering the limitations of our dataset, but improvements are needed to make the model more practically viable for scientific research.

\section{Interpretability}
MIL models provide researchers with increased interpretability of the model's predictions and localization, without finely resolved labels \cite{early2024inherently, Zhu2021}. The gated attention mechanism provides an insight into the contribution of each frame to the overall shot classification \cite{Ilse2018}. In this section, we provide some illustrative examples of the interpretability of our MIL model. We retrained a MIL model with the same configuration as discussed earlier, producing figures only from the 173 shots in the held-out test set.

\begin{figure*}[t]
\centering
\setlength{\figurewidth}{\minof{17cm}{\linewidth}}
\begin{minipage}{\figurewidth}
\centering
\begin{minipage}[t]{0.49\linewidth}
\textbf{(a)}\par
\includegraphics[width=\linewidth]{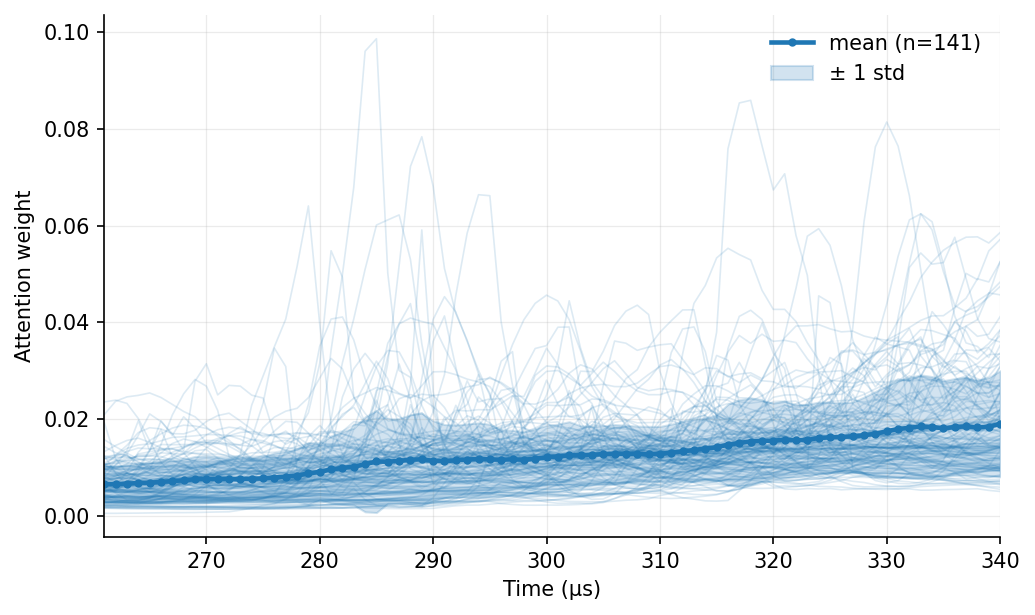}
\end{minipage}
\hfill
\begin{minipage}[t]{0.49\linewidth}
\textbf{(b)}\par
\includegraphics[width=\linewidth]{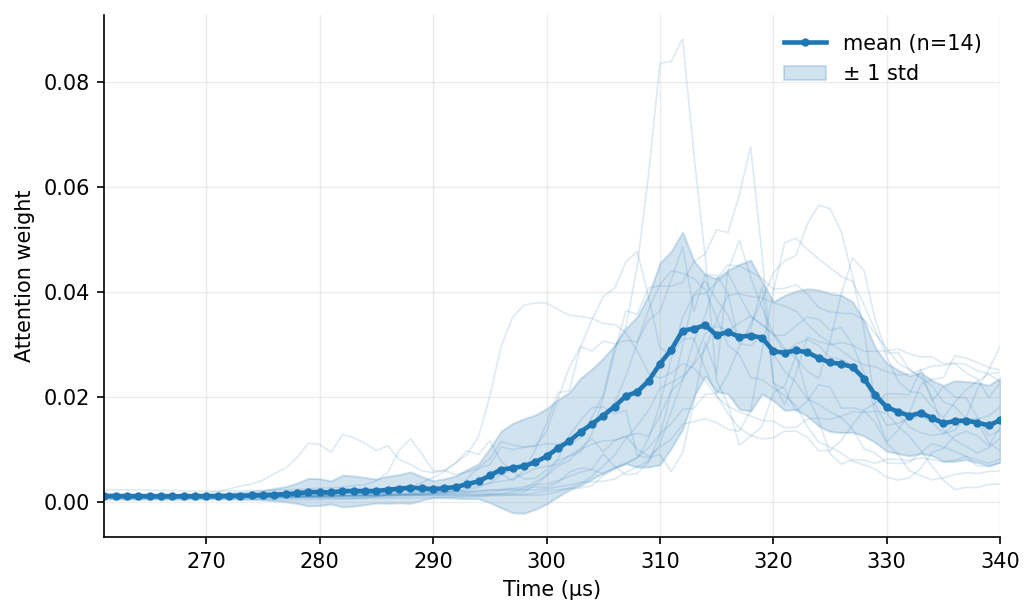}
\end{minipage}
\end{minipage}
\caption{Attention scores for every true negative (left) and true positive (right) shot. These plots were produced by overlaying the per-frame attention scores for each shot in these two categories. Non-plasmoid shots have no particularly high-weighted frames, whereas plasmoid shots typically have additional predictive information conveyed between 310-330 $\mu$s.}
\label{fig:combined_attention}
\end{figure*}

The attention mechanism can act as a tool for the temporal localization of plasmoids in plasmoid shots. In this section, we present a qualitative analysis of attention scores outputted by our MIL model. Figure \ref{fig:combined_attention} is an overlay of the attention plots for each four-channel frame of the 141 true negative (left) and 14 true positive (right) shots. Shots corresponding to true negative shot predictions tend to contain no frames of particular interest to the model, with a mostly flat, unchanging mean attention of approximately 0.01 across the entire range of frames. In the plot of the shots correctly classified as plasmoid shots, however, a distinct peak is apparent. Between frames 50--60, mean attention sits just above approximately 0.03, with an attention of $\geq 0.01$ in frames 40 onward. The model learns to weight these frames more strongly as they encode a greater amount of predictive information. A second important trend is that frames 0 to 30 have mean attention weights of approximately 0. This model learns to suppress such frames, as there is little meaningful information conveyed in these noisy and early-stage periods of each shot. It is important to note that high attention weights do not necessarily indicate a higher probability that a plasmoid is present in a certain frame; rather, they indicate the model assigns greater weight to that frame's learned representation when pooling all frames in a series.

\begin{figure*}[t]
\centering
\setlength{\figurewidth}{\minof{17cm}{\linewidth}}
\begin{minipage}{\figurewidth}
\centering
\begin{minipage}{0.75\linewidth}
\textbf{(a)}\par
\includegraphics[width=\linewidth]{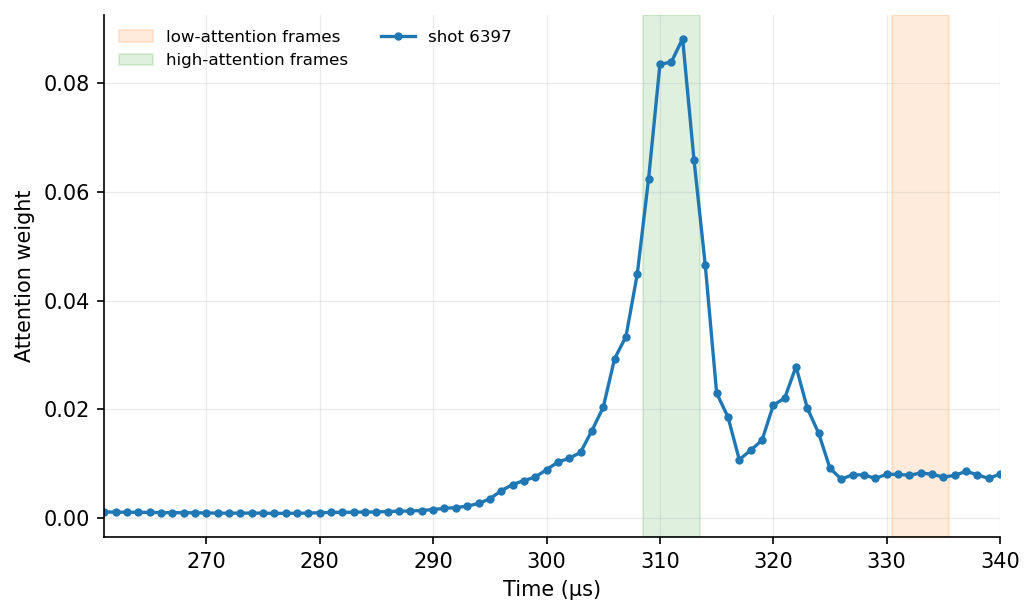}
\end{minipage}

\medskip

\begin{minipage}{\linewidth}
\textbf{(b)}\par
\includegraphics[width=\linewidth]{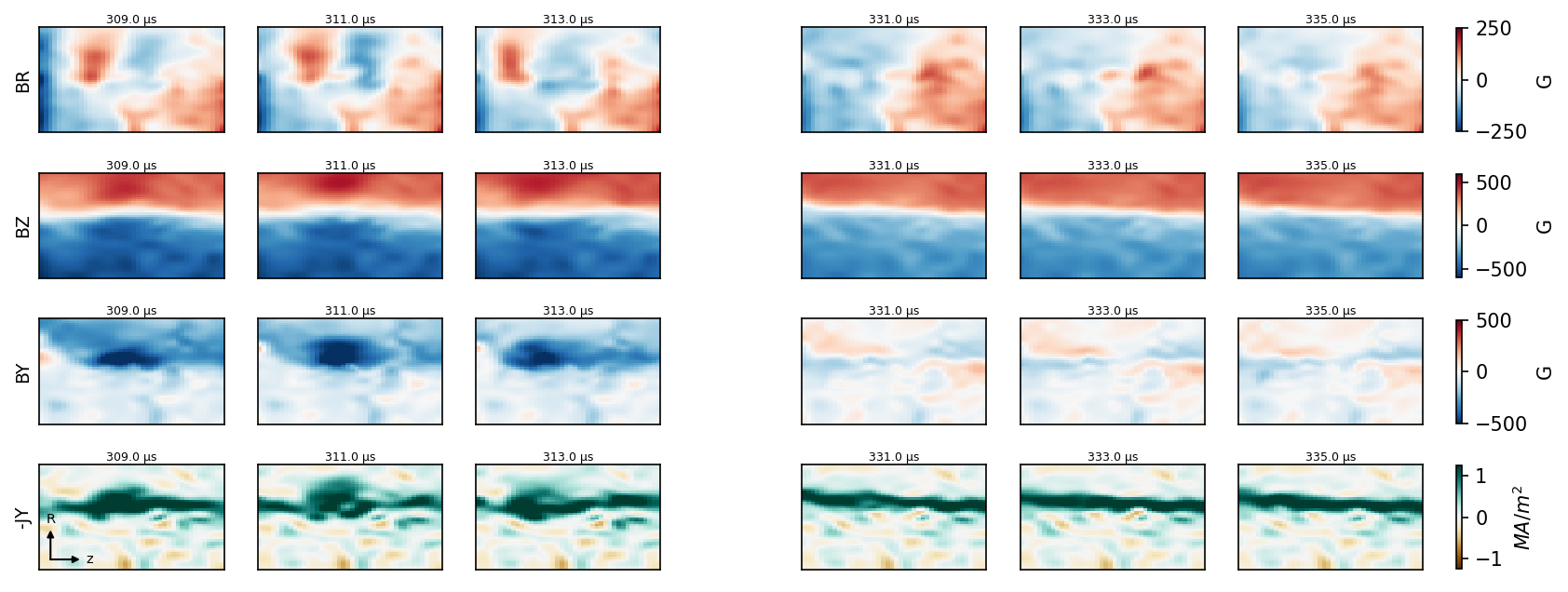}
\end{minipage}
\end{minipage}
\caption{Upper panel: Attention plot for shot 6397 between 261 $\mu$s and 340 $\mu$s, with frames at 1 $\mu$s intervals. Highlighted temporal regions correspond to the time intervals of high- and low-attention frames displayed in the lower panel. Lower panel: Plot of shot 6397 spatial profile frames in regions of high attention (left, 309 $\mu$s to 313 $\mu$s) and low attention (right, 331 $\mu$s to 335 $\mu$s). Each of the four BR, BZ, BY magnetic spatial profiles and the JY current density profile is displayed on separate rows with temporally aligned columns. Plasmoid-like structure is clearly apparent in the high-attention BY and JY frames, with distinct differences in BZ and BY data present in the two windows. Each frame is displayed with z horizontal, increasing to the right, and R vertical, increasing upwards.}
\label{fig:6397attention}
\label{fig:four_channel_highlow_comparison}
\end{figure*}

Let us examine a particular shot to understand the physical spatial profile features that correlate with high attention weights. Here, we choose shot 6397 for its remarkably strong and narrow main peak, contrasting the highest-attention frames with later, low-attention frames during which only the current sheet is present. See Figure \ref{fig:6397attention} for its attention plot. As in Figure \ref{fig:combined_attention}, frames from the first $\sim$30s effectively contribute no information to the series vector representation and the main peak aligns temporally with a large, rounded structure in the BY and JY channels.

We can also infer the presence of multiple plasmoids, in the case of shot 6397, from the bimodality of the attention curve. A manual inspection of spatial profile data confirmed that the second, smaller peak in frames 58--63 corresponds to a smaller plasmoid ejected in the opposite direction to the much larger first plasmoid. It could therefore be possible to discern the number of plasmoids in any shot by analyzing the modality of their attention weights.

\section{Conclusions}
In this paper, we have described a novel application of machine learning to plasma physics classification tasks. Specifically, we applied multiple instance learning to the problem of plasmoid classification using experimental time series spatial profile data. These classifiers are useful for studying plasmoids that occur in astrophysical systems, and subsequently better understanding reconnection, e.g., in the magnetosphere. Such weakly supervised methods are useful for when time-resolved labeling is prohibitively time consuming or otherwise infeasible. 

The MIL attention mechanism allows for inherent interpretability of model outputs. Through this mechanism, the model can learn to suppress the contribution of noisy frames in a plasmoid shot while assigning greater weight to later frames with plasmoid-like structure.
This is directly applicable to other plasma experiments where the data contain substantial noise, as gated attention can effectively highlight the scientifically useful components in a collection of data.
By plotting attention scores for frames in a time series, we could discern frames and times containing plasmoid-like structures. We also show how the modality of the attention curve can act attention scores could allow researchers to retroactively understand the causes of false predictions by examining the specific data that the model weights most heavily when aggregating information.

We found a statistically significant performance improvement relative to a convolutional recurrent neural network, a model widely used for plasma and fusion classification tasks involving time series data (see Table \ref{tab:wilcoxon_crnn_vs_gated_four_channel}). The MIL model also scored substantially higher in PR AUC, an important metric given the large class imbalance in our dataset (14.6\%, 5.9:1). However, the model is currently insufficient as an autonomous classifier. It achieved promising but not outstanding results on held-out test data. More work is needed to investigate preprocessing methods and develop the model into a tool widely used by researchers in similar areas of plasma physics.

Increasing the dataset size and reducing class imbalance should be prioritized. As FLARE is continuously used and upgraded over its lifespan, the size of the dataset will increase significantly, and we hope and expect a greater number of plasmoids will be found in the device's future shots. Future work will look to incorporate transfer learning to improve feature extraction. \citet{TimeMIL} extend MIL to account for time-ordering. Furthermore, the use of unsupervised clustering techniques on plasmoids detected with MIL methods would enable a more thorough understanding of the different types of plasmoids that form within experimental reconnection devices.



%
%

%


\begin{acknowledgments}
The authors acknowledge the DOE for their funding of the FLARE operation under contract \# DE-AC02-09CH11466. C.M. was supported by a University of Chicago Metcalf Grant. C.M. thanks Peiyun Shi for assisting in data acquisition and Akansha Bagga for insightful discussions of MIL. Any opinions, findings, and conclusions or recommendations expressed in this material are those of the authors and do not necessarily reflect the views of the funding organizations.
\end{acknowledgments}

\section*{Author Declarations}
\subsection*{Conflict of Interest}
The authors have no conflicts to disclose.

\subsection*{Author Contributions}
Conceptualization: C. Marsden, A. Robbins;
Data curation: C. Marsden, A. Robbins;
Formal analysis: C. Marsden;
Funding acquisition: C. Marsden, H. Ji;
Investigation: C. Marsden;
Methodology: C. Marsden;
Project administration: C. Marsden, A. Robbins, H. Ji;
Resources: C. Marsden, A. Robbins;
Software: C. Marsden;
Supervision: A. Robbins, H. Ji;
Validation: C. Marsden;
Visualization: C. Marsden;
Writing -- original draft: C. Marsden;
Writing -- review \& editing: C. Marsden, A. Robbins, H. Ji.

\section*{Data Availability}
The code used for this work can be found at \url{https://github.com/charlesmarsden/MIL-Plasmoid-Classification}. BR, BY, BZ, and JY spatial profile data can be found at Ref.~\onlinecite{marsden2026zenodo}. All machine learning model construction and training was done in PyTorch version 2.12.1, available at \url{https://pytorch.org} under a BSD-3-Clause License. Statistical tests were done using SciPy version 1.18.0, available at \url{https://scipy.org}, BSD-3-Clause License. Array operations were done using NumPy version 2.5.1, available at \url{https://numpy.org} under a BSD-3-Clause License. Figures were constructed using Matplotlib version 3.11.0, available at \url{https://matplotlib.org} under a Matplotlib License. All work was done using Python version 3.12.13.

\bibliographystyle{aipnum_doi}
\bibliography{galaxy}

@ARTICLE{gerhardt2008,
       author = {{Gerhardt}, S. P. and {Belova}, E. V. and {Yamada}, M. and {Ji}, H. and {Ren}, Y. and {McGeehan}, B. and {Inomoto}, M.},
        title = "{Field-reversed configuration formation scheme utilizing a spheromak and solenoid induction}",
      journal = {Physics of Plasmas},
         year = 2008,
        month = mar,
       volume = {15},
       number = {3},
          eid = {032503},
        pages = {032503},
          doi = {10.1063/1.2889428},
       adsurl = {https://ui.adsabs.harvard.edu/abs/2008PhPl...15c2503G}
}

@misc{ji2026flarefacility,
      title={The {FLARE} {Facility}}, 
      author={Ji, H. and Yoo, J. and Shi, P. and Jung, E. and Maheshwari, K. and Robbins, A. and Son, S. and Stanier, A. and Ren, Y. and Bose, S. and Corl, D. and Corrigan, K. and Cutler, R. and Daughton, W. and Ellis, R. and Gettelfinger, G. and Hatcher, R. and Heitzenroeder, P. and Hoffmann, F. and Jara-Almonte, J. and Kalish, M. and Kozub, T. and Merino, E. and Que, W. and Smith, B. and Wallace, J. and Zhao, X. and Avrutsky, S. and Crocker, N. and Dorfman, S. and Doke, Y. and Dunn, C. and Fox, W. and Li, B. and Li, X. and Myers, C. and Okazaki, S. and Pawlak, J. and Rhee, T. and Seo, B. and Suzuki, T. and Zhang, Y. and Bale, S. and Bhattacharjee, A. and Carter, T. and Dong, C. and Drake, J. and Egedal, J. and Gilson, E. and Inomoto, M. and Menard, J. and Ono, Y. and Prager, S. and Sarff, J. and Tanabe, H. and Yamada, M.},
      year={2026},
      eprint={2608.17332},
      archivePrefix={arXiv},
      primaryClass={physics.plasm-ph},
      url={https://arxiv.org/abs/2608.17332}, 
}

@article{Loureiro_2016,
doi = {10.1088/0741-3335/58/1/014021},
url = {https://doi.org/10.1088/0741-3335/58/1/014021},
year = {2015},
month = {nov},
publisher = {IOP Publishing},
volume = {58},
number = {1},
pages = {014021},
author = {Loureiro, N. F. and Uzdensky, D. A.},
title = {Magnetic reconnection: from the {Sweet–Parker} model to stochastic plasmoid chains},
journal = {Plasma Physics and Controlled Fusion}
}

@article{Loureiro2007,
    author = {Loureiro, N. F. and Schekochihin, A. A. and Cowley, S. C.},
    title = {Instability of current sheets and formation of plasmoid chains},
    journal = {Physics of Plasmas},
    volume = {14},
    number = {10},
    pages = {100703},
    year = {2007},
    month = {10},
    issn = {1070-664X},
    doi = {10.1063/1.2783986},
    url = {https://doi.org/10.1063/1.2783986},
    eprint = {https://pubs.aip.org/aip/pop/article-pdf/doi/10.1063/1.2783986/15762819/100703_1_online.pdf},
}

@article{JiPhaseDiagram,
    author = {Ji, H. and Daughton, W.},
    title = {Phase diagram for magnetic reconnection in heliophysical, astrophysical, and laboratory plasmas},
    journal = {Physics of Plasmas},
    volume = {18},
    number = {11},
    pages = {111207},
    year = {2011},
    month = {10},
    issn = {1070-664X},
    doi = {10.1063/1.3647505},
    url = {https://doi.org/10.1063/1.3647505},
    eprint = {https://pubs.aip.org/aip/pop/article-pdf/doi/10.1063/1.3647505/14910995/111207_1_online.pdf},
}

@article{KatesHarbeck2019,
  author = {Kates-Harbeck, J. and Svyatkovskiy, A. and Tang, W.},
  title = {Predicting disruptive instabilities in controlled fusion plasmas through deep learning},
  journal = {Nature},
  year = {2019},
  volume = {568},
  number = {7753},
  pages = {526--531},
  doi = {10.1038/s41586-019-1116-4},
  url = {https://doi.org/10.1038/s41586-019-1116-4},
  isbn = {1476-4687}
}

@article{Matos2020,
  author  = {Matos, F. and Menkovski, V. and Felici, F. and Pau, A. and Jenko, F. and {the TCV Team} and {the EUROfusion MST1 Team}},
  title   = {Classification of tokamak plasma confinement states with convolutional recurrent neural networks},
  journal = {Nuclear Fusion},
  year    = {2020},
  volume  = {60},
  number  = {3},
  pages   = {036022},
  doi     = {https://doi.org/10.1088/1741-4326/ab6c7a}
}

@article{Matos2021,
  author  = {Matos, F. and Menkovski, V. and Pau, A. and Marceca, G. and Jenko, F. and {the TCV Team}},
  year    = {2021},
  title   = {Plasma confinement mode classification using a sequence-to-sequence neural network with attention},
  journal = {Nuclear Fusion},
  volume  = {61},
  number  = {4},
  pages   = {046019},
  doi     = {https://doi.org/10.1088/1741-4326/abe370}
}

@article{Montes2021,
  author  = {Montes, K. J. and Rea, C. and Tinguely, R. A. and Sweeney, R. and Zhu, J. and Granetz, R. S.},
  title   = {A semi-supervised machine learning detector for physics events in tokamak discharges},
  journal = {Nuclear Fusion},
  year    = {2021},
  volume  = {61},
  number  = {2},
  pages   = {026022},
  doi     = {https://doi.org/10.1088/1741-4326/abcdb9}
}

@article{Zheng2020,
  author  = {Zheng, W. and Wu, Q. Q. and Zhang, M. and Chen, Z. Y. and Shang, Y. X. and Fan, J. N. and Pan, Y. and {J-TEXT Team}},
  title   = {Disruption predictor based on neural network and anomaly detection on {J-TEXT}},
  journal = {Plasma Physics and Controlled Fusion},
  year    = {2020},
  volume  = {62},
  number  = {4},
  pages   = {045012},
  doi     = {https://doi.org/10.1088/1361-6587/ab6b02}
}

@article{Churchill2020,
  author  = {Churchill, R. M. and Tobias, B. and Zhu, Y. and {the DIII-D Team}},
  title   = {Deep convolutional neural networks for multi-scale time-series classification and application to tokamak disruption prediction using raw, high temporal resolution diagnostic data},
  journal = {Physics of Plasmas},
  year    = {2020},
  volume  = {27},
  number  = {6},
  pages   = {062510},
  doi     = {https://doi.org/10.1063/1.5144458}
}

@article{Bergstedt2024,
  author  = {Bergstedt, K. and Ji, H.},
  title   = {A Novel Method to Train Classification Models for Structure Detection in In Situ Spacecraft Data},
  journal = {Earth and Space Science},
  year    = {2024},
  volume  = {11},
  number  = {6},
  pages   = {e2023EA002965},
  doi     = {https://doi.org/10.1029/2023EA002965}
}

@article{Dietterich1997,
  author  = {Dietterich, T. G. and Lathrop, R. H. and Lozano-P{\'e}rez, T.},
  title   = {Solving the multiple instance problem with axis-parallel rectangles},
  journal = {Artificial Intelligence},
  year    = {1997},
  volume  = {89},
  number  = {1--2},
  pages   = {31--71},
  doi     = {https://doi.org/10.1016/S0004-3702(96)00034-3}
}

@article{Doran2016,
  author  = {Doran, G. and Ray, S.},
  title   = {Multiple-Instance Learning from Distributions},
  journal = {Journal of Machine Learning Research},
  year    = {2016},
  volume  = {17},
  number  = {128},
  pages   = {1--50},
  url     = {http://jmlr.org/papers/v17/15-171.html}
}

@inproceedings{Ilse2018,
  author    = {Ilse, M. and Tomczak, J. M. and Welling, M.},
  title     = {Attention-based Deep Multiple Instance Learning},
  booktitle = {{Proceedings of the 35th International Conference on Machine Learning}},
  year      = {2018},
  volume    = {80},
  series    = {Proceedings of Machine Learning Research},
  pages     = {2127--2136},
  publisher = {PMLR},
  url       = {https://proceedings.mlr.press/v80/ilse18a.html}
}

@article{Schmidt2019,
  author        = {Schmidt, R. M.},
  title         = {Recurrent Neural Networks ({RNNs}): A gentle Introduction and Overview},
  year          = {2019},
  journal       = {arXiv preprint arXiv:1912.05911},
  doi           = {https://doi.org/10.48550/arXiv.1912.05911},
  url           = {https://arxiv.org/abs/1912.05911}
}

@article{LeCun1998,
  author  = {LeCun, Y. and Bottou, L. and Bengio, Y. and Haffner, P.},
  title   = {Gradient-Based Learning Applied to Document Recognition},
  journal = {Proceedings of the IEEE},
  year    = {1998},
  volume  = {86},
  number  = {11},
  pages   = {2278--2324},
  doi     = {https://doi.org/10.1109/5.726791}
}

@article{Deng2012,
  author  = {Deng, L.},
  title   = {The {MNIST} Database of Handwritten Digit Images for Machine Learning Research [Best of the Web]},
  journal = {IEEE Signal Processing Magazine},
  year    = {2012},
  volume  = {29},
  number  = {6},
  pages   = {141--142},
  doi     = {https://doi.org/10.1109/MSP.2012.2211477}
}

@inproceedings{
early2024inherently,
title={Inherently Interpretable Time Series Classification via Multiple Instance Learning},
author={Early, J. and Cheung, G. and Cutajar, K. and Xie, H. and Kandola, J. and Twomey, N.},
booktitle={{The Twelfth International Conference on Learning Representations}},
year={2024},
url={https://openreview.net/forum?id=xriGRsoAza}
}

@inproceedings{Zhu2021,
  author    = {Zhu, Y. and Shi, W. and Pandey, D. S. and Liu, Y. and Que, X. and Krutz, D. E. and Yu, Q.},
  title     = {Uncertainty-Aware Multiple Instance Learning from Large-Scale Long Time Series Data},
  booktitle = {2021 {IEEE} {International} {Conference} on {Big Data} ({Big Data})},
  year      = {2021},
  pages     = {1772--1778},
  doi       = {https://doi.org/10.1109/BigData52589.2021.9671469}
}

@misc{marsden2026zenodo,
  author       = {Marsden, C. and Robbins, A. and Ji, H.},
  title        = {{BR}, {BZ}, {BY}, {JY} spatial profile and shot label data},
  year         = {2026},
  howpublished = {Zenodo},
  note         = {Dataset},
  doi          = {https://doi.org/10.5281/zenodo.22048487}
}

@InProceedings{TimeMIL,
  title = 	 {{T}ime{MIL}: Advancing Multivariate Time Series Classification via a Time-aware Multiple Instance Learning},
  author =       {Chen, X. and Qiu, P. and Zhu, W. and Li, H. and Wang, H. and Sotiras, A. and Wang, Y. and Razi, A.},
  booktitle = 	 {{Proceedings of the 41st International Conference on Machine Learning}},
  pages = 	 {7190--7206},
  year = 	 {2024},
  editor = 	 {Salakhutdinov, R. and Kolter, Z. and Heller, K. and Weller, A. and Oliver, N. and Scarlett, J. and Berkenkamp, F.},
  volume = 	 {235},
  series = 	 {Proceedings of Machine Learning Research},
  month = 	 {21--27 Jul},
  publisher =    {PMLR},
  url = 	 {https://proceedings.mlr.press/v235/chen24af.html}
}

\end{document}